%% file: main.tex
\documentclass[%
    reprint, 
    superscriptaddress,
    amsmath,amssymb,
    aps,
    pra,
    floatfix,
]{revtex4-2}

\input{pfafflab-draft-header.tex}

\begin{document}

\title{Loss resilience of driven-dissipative remote entanglement in chiral waveguide quantum electrodynamics}

\author{Abdullah Irfan}
\affiliation{Department of Physics, University of Illinois Urbana-Champaign, Urbana, IL 61801, USA}
\author{Mingxing Yao}
\affiliation{Pritzker School of Molecular Engineering, University of Chicago, Chicago, IL 60637, USA}
\author{Andrew Lingenfelter}
\affiliation{Pritzker School of Molecular Engineering, University of Chicago, Chicago, IL 60637, USA}
\affiliation{Department of Physics, University of Chicago, Chicago, IL 60637, USA}
\author{Xi Cao}
\affiliation{Department of Physics, University of Illinois Urbana-Champaign, Urbana, IL 61801, USA}
\author{Aashish A. Clerk}
\affiliation{Pritzker School of Molecular Engineering, University of Chicago, Chicago, IL 60637, USA}
\author{Wolfgang Pfaff}
\affiliation{Department of Physics, University of Illinois Urbana-Champaign, Urbana, IL 61801, USA}
\affiliation{Materials Research Laboratory, University of Illinois Urbana-Champaign, Urbana, IL 61801, USA}



\begin{abstract}
    Establishing limits of entanglement in open quantum systems is a problem of fundamental interest, with strong implications for applications in quantum information science.
    Here, we study limits of entanglement stabilization between remote qubits.
    We theoretically investigate the loss resilience of driven-dissipative entanglement between remote qubits coupled to a chiral waveguide. 
    We find that by coupling a pair of storage qubits to the two driven qubits, the steady state can be tailored such that the storage qubits show a degree of entanglement that is higher than what can be achieved with only two driven qubits coupled to the waveguide. 
    By reducing the degree of entanglement of the driven qubits, we show that the entanglement between the storage qubits becomes more resilient to waveguide loss. 
    Our analytical and numerical results offer insights into how waveguide loss limits the degree of entanglement in this driven-dissipative system, and offers important guidance for remote entanglement stabilization in the laboratory, for example using superconducting circuits.
\end{abstract}

\maketitle



\section{Introduction}
\label{sec:intro}

Quantum reservoir engineering is a powerful paradigm to make use of the environment of a system to engineer or stabilize its quantum state \cite{poyatosQuantumReservoirEngineering1996}.
With entanglement being one of the defining features of quantum mechanics, it is particularly interesting to understand the conditions under which spatially distributed entangled states can be stabilized. 
Protocols for stabilizing entanglement are typically based on engineering the coupling between the qubits and a shared lossy environment such that the collective dissipation of the qubits relaxes them into an entangled state \cite{goviaStabilizingTwoqubitEntanglement2022, maStabilizingBellStates2019, maCouplingmodulationMediatedGeneration2021, stannigelDrivendissipativePreparationEntangled2012, motzoiBackactiondrivenRobustSteadystate2016, brownTradeOfffreeEntanglement2022}. Entanglement stabilization has also been experimentally demonstrated on multiple physical platforms including atoms in cavities \cite{kastoryanoDissipativePreparationEntanglement2011}, trapped ions \cite{coleResourceEfficientDissipativeEntanglement2022}, and superconducting circuits \cite{brownTradeOfffreeEntanglement2022, shankarAutonomouslyStabilizedEntanglement2013, kimchi-schwartzStabilizingEntanglementSymmetrySelective2016, shahStabilizingRemoteEntanglement2024}. 

\begin{figure}[!h]

    \centering
    \includegraphics{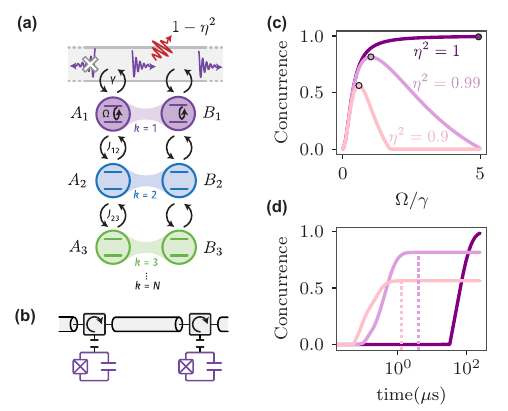}
    \caption{
        \textbf{(a)} A schematic of two qubit chains with the boundary qubits coupled to a unidirectional waveguide. 
        The boundary qubits are driven by a common Rabi drive with strength $\Omega$. 
        $J_{i,i+1}$ is the rate at which neighboring qubits in a chain exchange excitations. 
        Wave guide loss is modeled using a fictitious beam-splitter which allows a photon to pass through with probability $\eta^2$. 
        \textbf{(b)} Sketch of a possible circuit quantum electrodynamics implementation of the schematic in (a) with $N=1$. 
        \textbf{(c)} Steady state concurrence of the system with $N=1$ for different degrees of waveguide loss. See Section~\ref{sec:wg-loss} for details. \textbf{(d)} Time evolution of the concurrence for the ratios $\Omega/\gamma$ marked by circles in (c). The dotted vertical lines show the effective onsite decay time on the upstream qubit due to waveguide loss.
        }
    \label{fig:1}
\end{figure}

Here, we are particularly interested in entanglement stabilization between \textit{remote} qubits: qubits that are not coupled directly or even hybridized through a `bus' mode. 
Such a stabilization protocol---which is independent of the distance between the qubits---is also interesting from the perspective of quantum information processing. 
With the recent surge of interest in quantum networks, it has become important to be able to deterministically distribute entangled states between spatially separated qubits \cite{kimbleQuantumInternet2008, wehnerQuantumInternetVision2018, sangouardQuantumRepeatersBased2011}. 
In this context, we are interested in stabilizing and storing an entangled state distributed across two nodes of a network.
Our central aim is to understand how residual, non-engineered dissipation that is not a part of the stabilization scheme impacts stabilization performance and how it may be possible to overcome the resulting limitations.

We take as the starting point of our investigation a well understood result: 
two qubits dissipatively coupled to a 1D chiral waveguide can be driven into an entangled steady state \cite{stannigelDrivendissipativePreparationEntangled2012, motzoiBackactiondrivenRobustSteadystate2016}. 
Furthermore, it has been recently shown in Ref.~\cite{lingenfelterExactResultsBoundarydriven2023a} that if we replace the two qubits coupled to the chiral waveguide with chains of $N$ qubits, the system can relax into a highly entangled state that approaches a product of Bell pairs, as illustrated in Fig.~\ref{fig:1}.
This serves as a basis for stabilizing remote entanglement, and also storing the entangled state by preventing it from directly decaying into the waveguide.

Surprisingly, we find that by passively coupling a pair of storage qubits to the original system of two qubits coupled to the waveguide, not only do we stabilize entanglement in the storage pair, but the entanglement of the storage pair can exceed the maximum entanglement achievable in the original protocol, given any degree of photon loss in the waveguide. While the increase in the degree of entanglement is not huge, it is intriguing in itself to wonder why the storage qubits are more loss resilient compared to the driven qubits. We find that the photon loss in the waveguide effectively acts as a relaxation process on the upstream driven qubit. Remarkably, the exact solution found in Ref.~\cite{lingenfelterExactResultsBoundarydriven2023a} has a parameter regime that minimizes the population of the driven qubits, thus mitigating the effect of this induced relaxation process, and yet stabilizes a Bell pair on the storage qubits.
This results in an increase in the degree of entanglement of the storage qubits (compared with the original protocol) at the expense of the driven qubits. Our findings are complemented by an analytical understanding of the system in this regime: 
by adiabatically eliminating the driven qubits, we find that the effective drives on the storage qubits have a `built-in' asymmetry that counters waveguide loss, making them more loss resilient compared to the protocol with only two driven qubits. Finally, we find that due to the interplay between the driving, dissipation, and qubit-qubit couplings, there is an experimentally feasible parameter regime in which the storage qubit entanglement always exceeds the best entanglement of the original two-qubit protocol.

This paper is organized as follows. In Section~\ref{sec:2qubit}, we review how two qubits coupled to a 1D chiral waveguide relax into an entangled state, and provide an explanation for how waveguide loss limits the degree of entanglement of the steady state. In Section~\ref{sec:4qubit}, we discuss how adding a pair of storage qubits leads to a slightly improved resilience of the entangled state against waveguide loss. We explain the mechanism behind this improvement and quantify the gain in entanglement over the maximum entanglement achievable with only two driven qubits. Section~\ref{sec:morequbits} discusses the prospects of adding more qubits to each node and potential limits on how robust the steady state can be against waveguide loss.

\section{Stabilizing Two-qubit Remote Entanglement}
\label{sec:2qubit}

\subsection{A driven cascaded network of qubits}
We begin by considering a pair of qubit nodes connected via a unidirectional waveguide. 
Each node consists of a chain of qubits exchange-coupled with strength $J_{i,i+1}$. 
The first qubit is radiatively coupled to the waveguide with strength $\gamma$, and Rabi driven with strength $\Omega$, as illustrated in Fig.~\ref{fig:1}(a). 
The $N=1$ limit of this system, consisting of two driven qubits coupled to a unidirectional waveguide, has been studied previously and has been shown to have a steady state that can be highly entangled \cite{stannigelDrivendissipativePreparationEntangled2012, motzoiBackactiondrivenRobustSteadystate2016}. 
We briefly revisit this result for completeness, and then discuss how this system performs in the presence of waveguide loss.

The cascaded network of qubits can be modeled by the system Hamiltonian consisting of the qubits, the waveguide, and their interaction. 
Using a Markov approximation and tracing out the waveguide modes, we can obtain the equation of motion of the reduced two-qubit system, as originally described in \cite{gardinerDrivingQuantumSystem1993}. 
Alternatively, we can employ SLH formalism \cite{combesSLHFrameworkModeling2017} to identify an SLH triple (scattering matrix, Lindbladian, Hamiltonian) for each element in the network, and then use series composition rules to obtain the effective SLH triple for the cascaded system (see Appendix~\ref{app:slh} for details).
The master equation for a cascaded system of two driven qubits can then be written as
\begin{equation}
    \hat{\rho} = -i[\hat{H}, \hat{\rho}] + \gamma \mathcal{D}[\hat{c}]\hat{\rho} \label{eq:qme}
\end{equation}
where $\gamma$ is the coupling strength between the qubits and the waveguide. The waveguide-coupled Hamiltonian is given by
\begin{equation}
    \hat{H} = \frac{\Omega}{2}(\hat{\sigma}^{x}_A + \hat{\sigma}^{x}_B) + \frac{\Delta}{2}(\hat{\sigma}^{z}_A - \hat{\sigma}^{z}_B) + i\frac{\gamma}{2}(\hat{\sigma}^{+}_A \hat{\sigma}^{-}_B - \mathrm{h.c.}),
    \label{eq:2qHamiltonian-ideal}
\end{equation}
and the joint collapse operator is given by
\begin{equation}
    \hat{c} = \hat{\sigma}^{-}_A + \hat{\sigma}^{-}_B.
    \label{eq:collectiveLossIdeal}
\end{equation}
The Hamiltonian is written in the rotating frame of the common Rabi drive which has strength $\Omega$, and is detuned by $\Delta$ ($-\Delta$) from qubit A (B). 
The last term in Eq.~\eqref{eq:2qHamiltonian-ideal} describes the waveguide-mediated interaction between the qubits.
The joint collapse operator $\hat{c}$ arises from the interference between photons emitted from the two qubits. 
To find a pure state of this system, we look for a dark state (a state that gives zero when the collapse operator is applied to it), which describes a state of the system in which no photons propagate beyond the two qubits. 
One can see \cite{stannigelDrivendissipativePreparationEntangled2012, motzoiBackactiondrivenRobustSteadystate2016, pichlerQuantumOpticsChiral2015} that there is a dark state of this system that is also an eigenstate of the Hamiltonian with a zero eigenvalue, and therefore a steady state of the system, given by
\begin{equation}
    \ket{\psi_0} = \ket{00} + \frac{\sqrt{2}\Omega}{2\Delta - i\gamma}\ket{S},
\end{equation}
up to a normalization constant. 
$\ket{S}=(|01\rangle-|10\rangle)/\sqrt{2}$ is the singlet state. 
For $\Omega^2 \gg \Delta^2 + \gamma^2/4$, the dark steady state is entangled. 
The unidirectional nature of the waveguide gives rise to steady-state entanglement that is independent of the physical distance between the qubits.
From an application perspective, this protocol is thus highly interesting for realizing on-demand entanglement between distant qubits.
With superconducting qubits, for example, one could envision distributing entanglement across `modules' of a quantum processor \cite{burkhartErrorDetectedStateTransfer2021, leungDeterministicBidirectionalCommunication2019a, niuLowlossInterconnectsModular2023, campagne-ibarcqDeterministicRemoteEntanglement2018, axlineOndemandQuantumState2018, kurpiersDeterministicQuantumState2018}(Fig.~\ref{fig:1}(b)).
To assess the practical potential of this protocol, however, we must first understand how waveguide loss affects the attainable degree of entanglement.
In the following, our goal is to gain an intuitive picture of the effect of waveguide loss, as well as develop approaches to improve resilience against this loss.

\subsection{Understanding the effect of waveguide loss}
\label{sec:wg-loss}
We begin by deriving the collapse operators for our system in the presence of waveguide loss. 
We again use the SLH formalism for our cascaded system. 
A fictitious beam splitter is introduced in between the two qubits, described by the probability $\eta^2$ that the photon gets transmitted from A to B (see Appendix~\ref{app:slh} for details).
This leads to a multiplicative factor of $\eta$ in the waveguide-mediated qubit-qubit coupling term in the Hamiltonian, giving 
\begin{equation}
    \hat{H} = \frac{\Omega}{2}(\hat{\sigma}^{x}_A + \hat{\sigma}^{x}_B) + \frac{\Delta}{2}(\hat{\sigma}^{z}_A -\hat{\sigma}^{z}_B) + i\eta\frac{\gamma}{2}(\hat{\sigma}^{+}_A \hat{\sigma}^{-}_B - \mathrm{h.c.}).
    \label{eq:2qHamiltonian} 
\end{equation}
More interestingly, the collective loss operator becomes asymmetric in qubit-waveguide coupling, and a second loss operator on the upstream qubit is introduced, given by
\begin{align}
    \hat{c}_1 &= \eta \hat{\sigma}^{-}_A + \hat{\sigma}^{-}_B \nonumber \\
    \hat{c}_2 &= \sqrt{1-\eta^2} \hat{\sigma}^{-}_A. \label{eq:c2}
\end{align}
The asymmetry in the collective loss operator $\hat{c}_1$ is not automatically detrimental -- in principle, it could be countered simply by changing one of the qubit-waveguide coupling rates to restore `matching'.
That leaves $\hat{c}_2$ as the detrimental effect of waveguide loss on this system. 
The loss operator $\hat{c}_2$ describes losing a photon during propagation from qubit A to qubit B, which can be interpreted effectively as on-site loss on qubit A. 

For the remainder of the paper, we assume for simplicity that the drive detuning $\Delta=0$; nothing essential is lost by making this assumption. The effect of this effective on-site loss manifests as a reduction of the maximum concurrence, as shown in Figure~\ref{fig:1}(c). For each value of waveguide loss, we simulate the Lindblad master equation as a function of the ratio $\Omega/\gamma$ and compute the concurrence of the steady state. We see that for each $\eta^2$, the concurrence initially increases with $\Omega/\gamma$ up to a maximum, after which it begins to fall. 
To understand why increasing $\Omega/\gamma$ beyond this point causes a reduction in the concurrence, we consider the rates involved in this dissipative system.
For each value of $\eta^2$, we extract $\Omega/\gamma$ at the maximum and plot the time evolution of the concurrence for those parameters in Fig.~\ref{fig:1}(d). 
First, the system has a characteristic relaxation time which increases with $\Omega/\gamma$. This can be explained by the Liouvillian gap of the system which characterizes the slowest relaxation rate in the Liouvillian spectrum, and thus its inverse characterizes the relaxation timescale $\tau_{\rm rel}$. In this system, the relaxation times scales as $\tau_{\rm rel} \sim (\Omega/\gamma)^{2}$, i.e., the relaxation timescale grows quadratically with drive strength \cite{goviaStabilizingTwoqubitEntanglement2022, brownTradeOfffreeEntanglement2022}. As a result, there is a trade-off between the relaxation time and the degree of stabilized entanglement.
Second, we can see from Eq.~\ref{eq:c2} that the effective on-site loss on the upstream qubit introduces a decay rate given by $\gamma_{\rm loss} = \gamma \sqrt{(1-\eta^2)}$, thus limiting the $T_1$ of qubit $A_1$ to $T_1 < 1/\gamma_{\rm loss}$.

Heuristically, when the relaxation time exceeds this induced decay, $\tau_{\rm rel} > T_1$, the system cannot effectively stabilize greater entanglement because the upstream qubit decays too quickly.
For each $\eta^2$ in Fig.~\ref{fig:1}(d), the dotted lines mark the time $1/\gamma_{\rm loss} (\eta)$, which is the timescale associated with loss-induced decay. 
We see that these times qualitatively predict the onset of the concurrence plateau.
The effect of waveguide loss on this stabilization protocol can thus clearly not be cancelled entirely by adjusting coupling rates.
We discuss in Section~\ref{sec:4qubit} how the steady state can be made more loss resilient by adding additional qubits.

\subsection{Possible circuit QED implementation}
With an understanding of how loss affects the steady state concurrence of this system, we turn to a discussion of implementation using superconducting circuits. The most straightforward circuit implementation consists of transmon qubit modules connected through a unidirectional waveguide as illustrated in Fig.~\ref{fig:1}(b). A microwave circulator in the waveguide connecting the qubits ensures unidirectional photon propagation. The qubit-waveguide coupling $\gamma$ is determined by the strength of the capacitive coupling shown in Fig.~\ref{fig:1}(b). Similar circuits have been previously implemented for remote entanglement protocols that rely on the first qubit emitting a photon that propagates through a unidirectional waveguide and gets captured by the second qubit \cite{campagne-ibarcqDeterministicRemoteEntanglement2018, axlineOndemandQuantumState2018, kurpiersDeterministicQuantumState2018}. Based on the transmission loss reported in these papers, and more recent qubit module interconnect designs \cite{burkhartErrorDetectedStateTransfer2021}, we estimate that waveguide loss corresponding to $\eta^2 = 0.9$ is well within reach. Fig.~\ref{fig:1}(c) shows the steady state concurrence achievable with this amount of waveguide loss and the required ratio of drive strength and qubit-waveguide coupling.

\section{Protecting steady-state entanglement from waveguide loss}
\label{sec:4qubit}

\subsection{Coupling to storage qubits}

One possible idea for improving resilience is to add an additional pair of qubits:
For one, it is useful in general to be able to able to store the entangled state as a resource for further processing. 
Therefore it would make sense to transfer the entanglement to a pair of qubits that is not directly coupled to a loss channel. 
Further, envisioning the scheme as a resource for entanglement generation, one could further imagine that many storage qubits could be added to each node. 
In that direction it could be of interest to stabilize large entangled states using a single dissipative channel.
This could be achieved, for example, by propagating entanglement among qubits in a node using gates.

As a practical matter, particularly in the context of circuit QED, it is also important to consider the possible difficulty with reading out the qubits. In the case of this driven-dissipative protocol, one would turn off the Rabi drive before performing dispersive readout of the qubits, as is typically practiced in protocols involving driven transmons \cite{nguyenProgrammableHeisenbergInteractions2024, luMultipartiteEntanglementRabiDriven2022}. Since the qubits are coupled to the waveguide, the state would decay during the readout process. For ideal qubits that have no intrinsic relaxation, decay during the readout process could be minimized by reducing $\gamma$ while maintaining the required $\Omega/\gamma$ ratio. In an experiment, however, the qubits will have finite relaxation times which would compete with the decay rate into the waveguide given by $\gamma$. 
We show in Appendix~\ref{app:gamma_threshhold} that for qubits with a $T_1 = 100~\mu\mathrm{s}$ coupled to a chiral waveguide with $\eta^2=0.9$, the coupling to the waveguide $\gamma$ must be greater than $1~\mathrm{MHz}$ in order to achieve a concurrence of $0.56$. This would mean that when the drives are turned off, the qubit would decay in approximately $160~\mathrm{ns}$. 
We note that recently demonstrated fast high-fidelity readout times are comparable to this decay time \cite{sunadaFastReadoutReset2022, heinsooRapidHighfidelityMultiplexed2018}, hence preventing high-fidelity, single-shot readout of the quantum state. 
It is therefore also practically useful to be able to measure the entangled state in a pair of storage qubits that are decoupled from the waveguide.

Ref.~\cite{lingenfelterExactResultsBoundarydriven2023a} has shown that a system of two identical chains of coupled qubits coupled to the first pair has a pure steady state that stabilizes entanglement between the chains for any nonzero driving. Therefore, by coupling storage qubits to the waveguide-coupled pair, the system automatically stabilizes entanglement in the storage qubits.
Remarkably, we find that when there is loss in the waveguide, adding a pair of storage qubits increases the system's resilience against waveguide loss. While the improvement is modest, it is of fundamental interest to understand why there should be \emph{any} increase in resilience. To that end, we give first a heuristic argument as to why we might expect an improvement, then we analytically show (in part) how the increased resilience arises, and we quantify the performance improvement. 
We also find, as a practical matter, that the regime of maximally increased performance is reasonably achievable experimentally; the dissipation, driving, and hopping strengths should be similar in magnitude, and the region of improvement does not require fine tuning.

We start from the two-qubit entanglement stabilization protocol outlined in Sec.~\ref{sec:2qubit}, with the ideal (i.e., lossless waveguide, $\eta=1$) Hamiltonian Eq.~\eqref{eq:2qHamiltonian-ideal} and collective jump operator Eq.~\eqref{eq:collectiveLossIdeal}. 
At each node $A$ ($B$), we add a second qubit that is exchange-coupled with strength $J_{12,A(B)}$ to the first qubit (see Fig.~\ref{fig:1}a).
The Hamiltonian of the full four-qubit system is
\begin{align}
    \hat H_{\rm 4qb} = \hat H_{\rm 2qb} + \sum_{s=A,B} J_{12,s}\left( \hat\sigma_{s,1}^+\hat\sigma_{s,2}^- + {\rm h.c.} \right), \label{eq:H4qb}
\end{align}
where $\hat H_{\rm 2qb}$ is the entanglement stabilization Hamiltonian Eq.~\eqref{eq:2qHamiltonian-ideal} (acting on qubits $A_1$ and $B_1$).
As mentioned earlier in the text, we assume for simplicity that the drive detuning $\Delta=0$. The first pair of qubits, $A_1$ and $B_1$, remain coupled to the waveguide through the collective loss dissipation Eq.~\eqref{eq:collectiveLossIdeal} (in the ideal waveguide limit).

\subsection{Steady state entanglement}

The four-qubit system is the $N=2$ case of the general $N{+}N$-qubit double chain studied in Ref.~\cite{lingenfelterExactResultsBoundarydriven2023a}. 
In that work, it was shown that if the exchange couplings have a mirror symmetry, $J_{i,i+1;A}=J_{i,i+1;B}$, then the entire double chain has a pure steady state for arbitrary parameters, with entanglement between the two chains for any driving $\Omega\neq 0$.
Following this, we assume that the exchange couplings are equal, $J_{12,A}=J_{12,B}\equiv J_{12}$. Therefore, the $2+2$-qubit system has the pure entangled steady state (up to normalization)
\begin{align}
    |\psi_2\rangle = \frac{2\Omega^2}{\gamma J_{12}}|S_1 T_2\rangle +\frac{\Omega}{\sqrt{2}J_{12}}|0_1 T_2\rangle - |0_1 0_2\rangle.
    \label{eq:2+2ss}
\end{align}
Here the states are: the singlet $|S_1\rangle=(|0_{A,1}1_{B,1}\rangle-|1_{A,1}0_{B,1}\rangle)/\sqrt{2}$ of qubits on site 1, the triplet $|T_2\rangle =(|0_{A,2}1_{B,2}\rangle+|1_{A,2}0_{B,2}\rangle)/\sqrt{2}$ on site 2, and the two-qubit vacuum $|0_j\rangle=|0_{A,j}0_{B,j}\rangle$ on either site $j$.

The steady state approaches a maximally entangled state $|\psi_2\rangle=|S_1 T_2\rangle$ when $\Omega^2 \gg \gamma J_{12}$, which can be achieved even if $\Omega < \gamma$ \cite{lingenfelterExactResultsBoundarydriven2023a}.
For the purpose of stabilizing and protecting entanglement, however, there is a related but different parameter regime of interest,
\begin{align}
    J_{12}\ll\Omega.
\end{align}
In this regime, the vacuum component $|0_1 0_2\rangle$ of the steady state is negligible so the state nearly factorizes as $|\psi_2\rangle\approx (2\Omega/\gamma |S_1\rangle+|0_1\rangle)\otimes |T_2\rangle$. Thus, the second qubit pair can be arbitrarily close to a perfect Bell pair, irrespective of the entanglement of the first pair.
By taking $\Omega \ll \gamma$ as well, with the hierarchy of scales $J_{12}\ll\Omega\ll\gamma$ and the driving strength satisfying $\Omega^2 \ll \gamma J_{12}$, entanglement is stabilized only on the second pair, $|\psi_2\rangle\approx |0_1 T_2\rangle$. Note that this is the $N=2$ case of the single ``charge-density-wave'' predicted to exist in these chains \cite{lingenfelterExactResultsBoundarydriven2023a}.

The ability to parametrically control the amount of entanglement stabilized on the second qubit pair independently of the entanglement on the first pair plays a crucial role in the increased resilience of the four-qubit system against loss in the waveguide. By using the first pair of qubits as a ``sacrificial" pair whose entanglement is intentionally made worse, we can improve the entanglement of the second pair beyond that which can be achieved for the two-qubit scheme for any waveguide transmission $\eta^2$ for which the two-qubit protocol can stabilize entanglement.

\subsection{Inherent resilience to waveguide loss}

We account for waveguide loss in the four qubit system exactly following Sec.~\ref{sec:wg-loss}. 
The collective loss dissipator Eq.~\eqref{eq:collectiveLossIdeal} is replaced by the two dissipators of Eq.~\eqref{eq:c2}, where $\eta$ is the waveguide transmission amplitude, and the dissipation-induced exchange term in $\hat{H}_{\rm 2qb}$ (cf.~Eq.~\eqref{eq:2qHamiltonian}) is modified as in Sec.~\ref{sec:wg-loss}.
Just as in the two-qubit system, there is no pure steady state for any $\eta<1$ due to the single qubit loss induced on qubit $A_1$.
However, notice that if the first pair of qubits is in vacuum, then the effective single qubit loss cannot disrupt the steady state; such a steady state would remain a dark state of the dissipation \footnote{It should be noted that because the collective loss dissipator is modified as well, the dark state condition is also broken, $\hat c_1|S_1\rangle\neq0$. 
The two-qubit vacuum is the unique dark state of the two dissipators.}.

\begin{figure}[tb]
    \centering
    \includegraphics{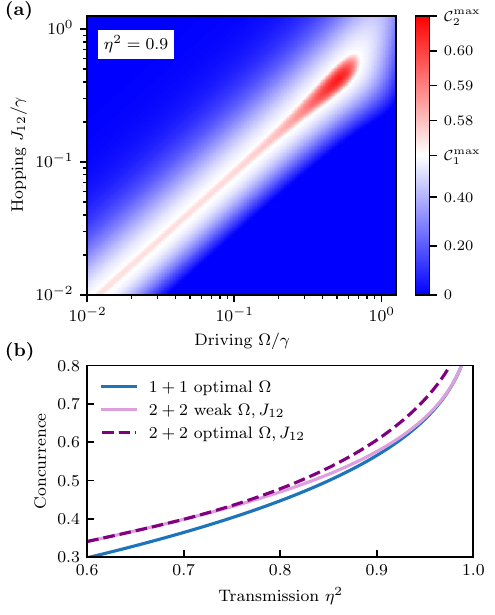}
    \caption{
        \textbf{(a)} Steady state concurrence of the outer pair of qubits in the $2+2$ system as a function of both driving $\Omega/\gamma$ and hopping $J_{12}/\Omega$. 
        Waveguide transmission probability is $\eta^2=90\%$. 
        The color scale is centered at the maximum concurrence of the $1+1$ system (with symmetric driving $\Omega_A = \Omega_B =\Omega$). 
        The red region indicates the range of $\Omega$ and $J_{12}$ for which the $2+2$ system stabilizes more entanglement than the $1+1$ system, $\mathcal{C}_1^{\rm max}\approx 0.57$, which is achieved for $\eta^2=0.9$ with a driving strength $\Omega/\gamma\approx 1.16$.
        Note that the color scale is not symmetric about $\mathcal{C}_1^{\rm max}$; the minimum is at zero concurrence and the maximum is at the maximum concurrence of the $2+2$ system for this waveguide loss, which is found numerically to be $\mathcal{C}_2^{\rm max}\approx 0.61$. 
        \textbf{(b)} Comparison of maximum concurrence for the $1+1$ system, optimizing over $\Omega$, and the maximum concurrence for the outer pair of the $2+2$ system in the weak $\Omega$, weak $J_{12}$ limit, optimizing over $J_{12}/\Omega$, and optimized over $\Omega$ and $J_{12}$.  
    }
    \label{fig:resilience}
\end{figure}

Guided by this insight, and recalling that the exact solution Eq.~\eqref{eq:2+2ss} approaches such a state with near zero population on the first qubit pair in the parameter regime set by $J_{12}\ll\Omega$ and $\Omega^2\ll \gamma J_{12}$, we are lead to the strategy of using the first pair of qubits as a sacrificial pair in order to stabilize entanglement on the second pair.
The utility of this strategy is verified numerically in Fig.~\ref{fig:resilience}(a), which shows that, as $J_{12},\Omega\to0$ (while holding $J_{12}/\Omega$ fixed), the concurrence on the second pair of qubits is non-vanishing, approaching a constant value dependent on the ratio $J_{12}/\Omega$. 
Moreover, there is a regime of constant $J_{12}/\Omega$ (with $\Omega>J_{12}$) for which the $2+2$ scheme yields higher concurrence for the $\eta^2$ used in the figure. 
A heuristic argument for the constant, non-vanishing concurrence with $J_{12},\Omega\to0$ is discussed in Appendix~\ref{app:const-conc}. 
We find numerically that for all transmission probabilities $\eta^2$ for which the $1+1$ system stabilizes entanglement, there is a parameter regime in which the $2+2$ system yields better concurrence than the $1+1$ system.

To better understand why the $2+2$ is more robust to waveguide loss than the $1+1$ system, we first consider the regime of weak hopping and driving, $J_{12},\Omega\ll \gamma$. 
In this regime, the first pair of qubits is almost completely in vacuum. Thus, we can adiabatically eliminate these qubits to find an effective master equation for the reduced density matrix of the second pair, $\hat{\rho}_2 = {\rm Tr}_{A1,B1}[\hat{\rho}]$, given by
\begin{align}
    \partial_t \hat{\rho}_2 &= -i[\hat{H}_{\rm eff},\hat{\rho}_2] + \sum_{j=1}^2 \mathcal{D}[\hat{L}_{j,{\rm eff}}] \hat{\rho}_2, \label{eq:effective-me}\\
    \hat{H}_{\rm eff} &= \frac{\Omega_{\rm eff}}{2} \Big[ \hat{\sigma}_{A,2}^x + (2\eta -1 ) \hat{\sigma}_{B,2}^x \Big]  \\
    &+ \frac{i\eta \gamma_{\rm eff}}{2}\Big(\hat{\sigma}^{+}_{A,2} \hat{\sigma}^{-}_{B,2} - {\rm h.c.}\Big),\nonumber\\
    \hat{L}_{1,{\rm eff}}& = \sqrt{\gamma_{\rm eff}} \Big(\eta \hat{\sigma}_{A,2}^- + \hat{\sigma}_{B,2}^-\Big), \\
    \hat{L}_{2,{\rm eff}} &= \sqrt{\gamma_{\rm eff}} \sqrt{1-\eta^2} \hat{\sigma}_{A,2}^-.
\end{align}
See Appendix~\ref{app:adiabatic} for details.
The effective master equation is precisely of the form of the two-qubit scheme with waveguide loss, but with asymmetric driving strengths on the two qubits. The renormalized drive strength is $\Omega_{\rm eff} = 2\Omega J_{12}/\gamma$, and dissipation strength is $\gamma_{\rm eff} = 4J_{12}^2/\gamma$. Notice that the waveguide transmission amplitude $\eta$ is \emph{not} renormalized.

The driving asymmetry of the effective master equation is $\eta$-dependent and recovers the ideal symmetric driving strengths for $\eta\to1$. 
Because $\eta$ is not renormalized, any improvement of the stabilized concurrence of the $2+2$ system in the $J_{12},\Omega\ll\gamma$ regime must be due to the asymmetry in the driving strength.
Indeed, we find in Fig.~\ref{fig:resilience}(b) that over a wide range of $\eta$, the effective asymmetric driving of the effective theory (in the weak $\Omega$, $J_{12}$ limit) yields higher concurrence than the symmetric driving (the $1+1$ system with optimal $\Omega$ in Fig.~\ref{fig:resilience}(b)). 
Intuitively, we expect that some degree of drive asymmetry should improve the stabilized entanglement because the waveguide loss induces  additional single qubit loss on the upstream qubit, thus a stronger drive -- relative to the drive on the downstream qubit -- is needed to compensate for the greater loss.

We also find that the effective master equation recovers the numerically observed result that as $J_{12},\Omega\to0$, the concurrence on the second qubit pair depends only on $\Omega/J_{12}$ (see Fig.~\ref{fig:resilience}(a)).
Since the concurrence of the $1+1$ system is controlled only by $\Omega_{\rm eff}/\gamma_{\rm eff}$, we find that in terms of the original system parameter, $\Omega_{\rm eff}/\gamma_{\rm eff} = \Omega/2J_{12}$, thus the concurrence is dependent only on the ratio $\Omega/J_{12}$ and not on their strengths relative to $\gamma$. A heuristic explanation for this behavior is given in Appendix~\ref{app:const-conc}.
Finally, we observe numerically that for $J_{12}\sim \Omega \lesssim \gamma$, there is a regime for which the concurrence is even higher compared to the $1+1$ concurrence. This effect is shown in Fig.~\ref{fig:resilience}(a) as the head of the red matchstick region. A full analytic understanding of this ``bump'' in concurrence remains an open question, but we speculate that it may be due to a kind of ``impedance matching'' among the driving, hopping, and dissipation dynamics. The improvement of the ``bump'' region over the weak driving regime is shown as the dashed curve in Fig.~\ref{fig:resilience}(b).

\subsection{Universal improvement over two-qubit protocol}

\begin{figure}[tb]
    \centering
    \includegraphics{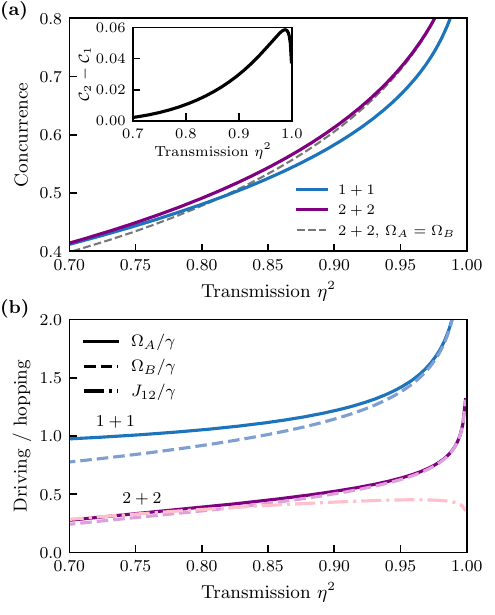}
    \caption{\textbf{(a)} Optimized concurrence of the $1+1$ system and the outer pair of the $2+2$ system as a function of transmission probability $\eta^2$. For each system we optimize over the upstream (downstream) Rabi drive $\Omega_A$ ($\Omega_B$) and for the $2+2$ system we optimize over $J_{12}$. For comparison, we also plot the best concurrence of the $2+2$ system with equal $\Omega_A=\Omega_B$ (cf. Fig.~\ref{fig:resilience}(b)). \textbf{Inset:} the difference in concurrence between the optimized $2+2$ system and the optimized $1+1$ system is plotted as a function of transmission probability. \textbf{(b)} The optimized drive strengths $\Omega_{A}$ / $\Omega_{B}$ of the upstream / downstream qubits and $J_{12}$ for the two systems plotted as functions of the transmission probability.
    }
    \label{fig:performance}
\end{figure}

As we have shown, the higher stabilized entanglement of the $2+2$ system compared with the $1+1$ system can be partially explained by the asymmetry of the effective Rabi drive strengths in the weak driving and weak hopping regime.
One may naturally wonder whether the $1+1$ system could achieve the same entanglement as the $2+2$ system if we optimized over the driven qubits' drives, $\Omega_A$ and $\Omega_B$, separately.
We find numerically that when allowing the applied driving strengths to have asymmetry, the $2+2$ system always has better concurrence on the outer pair of qubits than the $1+1$ system for a given $\eta^2$. 
In Fig.~\ref{fig:performance}(a) we show the maximum achievable concurrence of the $1+1$-qubit protocol, optimized over the driving strengths $\Omega_A,\Omega_B$, and the $2+2$-qubit protocol, optimized over $\Omega_{A}$, $\Omega_B$, and $J_{12}$, as a function of waveguide transmission probability $\eta^2$.
We find that the $2+2$ system stabilizes at least as much concurrence as the $1+1$ system for any $\eta^2>0$, including for $\eta^2$ not shown in the figure.
We also show the numerically optimized parameters $J_{12}$, $\Omega_A$, and $\Omega_B$ vs. waveguide transmission $\eta^2$ in Fig.~\ref{fig:performance}(b), and we find that except when the waveguide loss is very small ($\eta^2\approx 1$), the optimal parameters are $J_{12}\approx\Omega < \gamma$.
Furthermore, note that the asymmetry of the $2+2$ drives is relatively much smaller than the $1+1$ drives, suggesting that effective renormalization continues into the optimal parameter regime.
We emphasize that the improvement in stabilized concurrence of the $2+2$ system over the $1+1$ system is maximized for parameters $\Omega_{A,B}\approx J_{12}\approx\gamma$, and does not require one parameter to be much larger or much smaller than the others.
As a practical matter, this implies that the best performance can be obtained \emph{without} either precisely tuning one rate to be much smaller that the others, or engineering of one rate to be much stronger. (E.g., we do not need to work in the $\Omega_{A/B},J_{12} \ll \gamma$ regime.)

\section{Adding more qubits to each chain}
\label{sec:morequbits}

We have shown that by adding a second pair of qubits and sacrificing the driven pair, we can obtain greater stabilized entanglement in the face of waveguide loss. Given the exact solution for arbitrarily long chains found in Ref.~\cite{lingenfelterExactResultsBoundarydriven2023a}, a natural question might be whether there are advantages to coupling more qubits to the end of each chain. For example, does the resilience against waveguide loss improve by further isolating the entangled pair from the waveguide, e.g. by sacrificing the first $N-1$ pairs of an $N+N$-qubit system? Or, can we stabilize entanglement between many pairs of qubits within longer chains by sacrificing just the first pair, or a few pairs near the waveguide? Here we show that in principle, the intuition that guides us to sacrifice the first pair for $N=2$ can apply to longer chains. We also discuss the inherent challenges with adding more qubits to each chain.

\subsection{Stabilizing entanglement in longer chains}

We can add pairs of qubits to the ends of the chains exactly in the same manner as we added the second pair in Sec.~\ref{sec:4qubit}. We couple a qubit onto the end of each chain, maintaining equal hopping rates $J_{j,j+1;A}=J_{j,j+1;B}=J_{j,j+1}$. If this mirror symmetry is maintained, the exact solution of the ideal case found in Ref.~\cite{lingenfelterExactResultsBoundarydriven2023a} applies. The general form of the exact steady state is of a so-called ``hole pair  condensate'' in which the maximally entangled state $|STST\cdots\rangle$ is doped with pairs of adjacent holes (a hole on site $j$ being a two-qubit vacuum on qubits $A_j$ and $B_j$) placed along the chain. For any finite driving, $0<\Omega<\infty$, the chain is populated with all possible numbers of hole pairs. The disorder in the hopping rates $J_{j,j+1}$ affects the spatial distribution of the hole pair  wavefunctions. Specifically, the amplitude for a hole pair to span a strong bond is greater than average and the amplitude for a hole pair to span a weak bond is smaller. We can exploit this property of the hopping rates to keep the first pair of qubits near vacuum to counteract the effects of waveguide loss, as we did for the $N=2$ system in Sec.~\ref{sec:4qubit}.

Guided by the heuristic argument of Sec.~\ref{sec:4qubit}, we first show that in principle, large entanglement can be stabilized on qubit pairs $3$ through $N$ in a $N+N$ qubit system, while keeping the first pair near vacuum, by operating in a weak driving and weak hopping regime. The general form of the exact steady state for chains of length $N$ is
\begin{align}
    |\psi_N\rangle &= \left( 1+ \frac{i\gamma}{2\Omega}\hat\tau_1\right)\times\label{eq:weak-dr-param-regime}\\
    &\exp\left[\frac{i\gamma}{2\Omega^2}\sum_j (-1)^j J_{j,j+1}\hat\tau_j\hat\tau_{j+1}\right]|STST\cdots\rangle, \nonumber
\end{align}
where the operator $\hat\tau_j$ removes the Bell pair from the two qubits on site $j$: $\hat\tau_j|(S/T)_j\rangle = \sqrt2|0_j\rangle$ (see Ref.~\cite{lingenfelterExactResultsBoundarydriven2023a} for details). The exponential term describes the hole pair  condensate and there is a boundary correction that removes a single Bell pair from the dissipative qubit pair.
We find that if we work in the weak driving regime $\Omega\ll \gamma$, the Bell state occupation of the first pair of qubits is always suppressed by the boundary correction term. Then if we let the hopping rates be $J_{12}\gg J_{23} \approx J_{3,4} \approx \cdots \approx J_{N-1,N}$, the hole pairs are most strongly weighted on the first two pairs of qubits, leaving the rest of the chain relatively more entangled. Finally, notice that if $\Omega \gg \sqrt{\gamma J_{j,j+1}}$ for every $j>1$, then the hole pairs on those sites are suppressed. Thus we arrive at the parameter regime that generalizes the heuristic arguments from Sec.~\ref{sec:4qubit},
\begin{align}
    &\gamma \gg \Omega\approx J_{12} \gg J_{23}\approx J_{34}\approx\cdots,\\ &\Omega,J_{12} \gg \sqrt{\gamma J_{23}} \approx \sqrt{\gamma J_{34}} \approx \cdots.
\end{align}
We expect that this regime provides a starting point to optimize the entanglement in the rest of the chain in the presence of waveguide loss.
As an example, the ideal exact steady state of the $N=3$ system is given by
\begin{align}
    |\psi_3\rangle &= \left[|00S\rangle -\frac{\Omega}{\sqrt2J_{12}}|0TS\rangle \right] \\&-i \frac{\Omega^2}{\gamma J_{12}}|STS\rangle - \frac{J_{23}}{J_{12}}|S00\rangle -i\frac{J_{23}/J_{12}}{\Omega/\gamma} |000\rangle,\nonumber
\end{align}
up to normalization. In the parameter regime Eq.~\eqref{eq:weak-dr-param-regime}, the bracketed terms are dominant. Thus, we find that the third pair of qubits is highly entangled, while the second pair can be somewhat entangled depending on the ratio $\Omega/J_{12}$. All other terms are suppressed by the weak drive $\Omega\ll\gamma$ or the weak hopping $J_{23}\ll J_{12}$.  Note that the requirement for $\Omega,J_{12} \gg \sqrt{\gamma J_{j,j+1}}$ becomes apparent in e.g., the coefficient of the vacuum state $|000\rangle$.
For longer chains in the weak driving and weak hopping regime, one finds that there is a pair of terms equivalent to the bracketed pair above, given by $|00STST\cdots\rangle$ and $|0TSTST\cdots\rangle$. Thus by sacrificing the first pair, and to an extend the second pair, large entanglement can be stabilized along the rest of the chains.

\subsection{Potential limits on resilience against waveguide loss}

We found in Sec.~\ref{sec:4qubit} that the optimal concurrence of the second pair of qubits in the $N=2$ system occurs not deep in the weak driving limit but when $\Omega \approx J_{12} \lesssim \gamma$. We expect a similar result for longer chains, where the weak driving limit guides our intuition for choosing the relative strength of parameters, but ultimately, numerical optimization yields the best performance. It remains an open question, however, whether longer chains would continue to show improved resilience against waveguide loss compared to shorter chains.

One practical limiting factor is the relaxation time of the system. As discussed in Ref.~\cite{lingenfelterExactResultsBoundarydriven2023a}, there is numerical evidence to suggest the relaxation time of longer chains is consistent with the typical boundary driven free-fermion scaling with system size of $\tau_{\rm rel} \sim N^3$. Even for an ideal lossless waveguide, this poses an experimental challenge for longer chains due to the intrinsic loss and dephasing of the qubits. If the relaxation time becomes comparable to the typical qubit $T_1$ or $T_2$, these unwanted sources of dissipation will disrupt the entanglement stabilization. To counteract that, one must engineer stronger qubit-waveguide coupling $\gamma$ as well as stronger driving $\Omega$ and hopping rates $J_{j,j+1}$. The limits on the strength of the coupling and driving rates compared to intrinsic qubit dissipation rates thus limits the length of experimentally feasible chains.
It remains an open question what the ultimate performance of a $N+N$ system can be under waveguide loss and realistic intrinsic qubit dissipation.

\section{Conclusion}

We have proposed a driven-dissipative remote entanglement protocol and studied its resilience to waveguide loss. By understanding the loss mechanism that limits entanglement in the originally proposed scheme \cite{stannigelDrivendissipativePreparationEntangled2012}, and taking advantage of the pure steady state of the four-qubit system consisting of a driven pair and a storage pair, we propose a protocol that is more resilient to waveguide loss. 
Based on the intuition that waveguide loss can be countered by limiting the population on the driven qubits, we have identified an advantageous parameter regime characterized by weak drives and weak hopping between the driven and storage qubits. 
We find numerically that the degree of entanglement of the storage qubits in this regime is higher than the maximum possible entanglement that can be achieved with only two driven qubits. 
We explain this observation analytically by adiabatically eliminating the driven qubits to obtain an effective master equation for the storage qubits. 

While the driven-dissipative dynamics of the two-qubit system are well understood for a lossless waveguide \cite{stannigelDrivendissipativePreparationEntangled2012, motzoiBackactiondrivenRobustSteadystate2016}, our work provides an additional understanding of the dynamics in the presence of inevitable loss, and provides guidance on how to tailor operation parameters that yield better performance with photon loss. 
Our result is also interesting from a practical perspective, particularly in the context of circuit QED, where it may be difficult to faithfully measure qubits that are strongly coupled to a waveguide. 
Our work gives thus important guidance for practical implementation of driven-dissipative entanglement in the laboratory. 

\section*{Acknowledgements}
We acknowledge support from the National Science Foundation QLCI HQAN (NSF Award No. 2016136). 
AL, MY and AC acknowledge support from the Army Research Office under Grant No. W911NF-23-1-0077, and from the Simons Foundation through a Simons Investigator Award (Grant No. 669487).

\appendix

\section{Using the SLH formalism to derive the master equation}
\label{app:slh}

The SLH formalism provides a method to derive the Hamiltonian and collapse operators for a quantum network using simple algebraic manipulations \cite{combesSLHFrameworkModeling2017}. Each component is specified by a triple $(S, L, H)$ that describes how it interacts with input and output fields. $S$ is the scattering matrix for the component, $L$ contains its coupling to external fields, and $H$ is the Hamiltonian. The cascaded network of two qubits coupled to a unidirectional waveguide can be modeled using three components in series configuration: qubit A, followed by a fictitious beam splitter, followed by qubit B. The beam splitter has a probability $\eta^2$ of allowing a photon to propagate through from qubit A to qubit B. This allows us to model photon loss in the wave guide. The SLH triple for the $i$-th qubit is given by 
\begin{align*}
    S &= I \\
    L &= \sqrt{\gamma} \begin{pmatrix}
        \hat{\sigma}_{i}^- \\
        0
        \end{pmatrix} \\
    \hat{H} &= \frac{\Delta}{2} \hat{\sigma}_{i}^z + \frac{\Omega}{2} \hat{\sigma}_{i}^x.
\end{align*} 
The Hamiltonian is in the frame of the drive which is detuned from the qubit frequency by $\Delta$. The SLH triple for the beam splitter is given by
\begin{align*}
    S &= \begin{pmatrix}
        \eta & -\sqrt{1-\eta^2} \\
        \sqrt{1-\eta^2} & \eta
        \end{pmatrix} \\
    L &= 0 \\
    \hat{H} &= 0
\end{align*} where $\eta^2$ is the probability that a photon propagates through from qubit A to qubit B. 
For two components connected in series, the product rule used to calculate the SLH triple for the network is
\begin{align}
    &(S_2, L_2, H_2) \triangleleft (S_1, L1, H_1) = \label{eq:SLH_series}\\
    &\Big(S_2 S_1, L_2 + S_2 L1, H_1 + H_2 + \frac{1}{2i}(L_2^{\dagger} S_2 L_1 - L_1^{\dagger} S_2^{\dagger} L_2)\Big).\nonumber
\end{align}
Using Eq.~\ref{eq:SLH_series}, we obtain the following SLH triple for our cascaded system:

\begin{equation}
    \Bigg( 
            \begin{pmatrix}
            \sqrt{\eta} & \sqrt{1-\eta} \\
            -\sqrt{1-\eta} & \sqrt{\eta}
            \end{pmatrix},
            \sqrt{\gamma} \begin{pmatrix}
                \sqrt{1-\eta^2}\hat{\sigma}_{A}^{-} \\
                \eta \hat{\sigma}_{A}^{-} + \hat{\sigma}_{B}^{-}
                \end{pmatrix},
            \hat{H}
    \Bigg) 
\end{equation}
where the Hamiltonian is given by
\begin{align}
    \hat{H} &= \Big(\frac{\Delta_A}{2} \hat{\sigma}_{A}^{z} + \frac{\Omega_A}{2} \hat{\sigma}_{A}^{x} \Big) + \Big (\frac{\Delta_B}{2} \hat{\sigma}_{B}^{z} + \frac{\Omega_B}{2} \hat{\sigma}_{B}^{x} \Big) \\
    &- i \frac{\eta~\gamma}{2} \Big( \hat{\sigma}_{A}^{-}\hat{\sigma}_{B}^{+} - \hat{\sigma}_{A}^{+}\hat{\sigma}_{B}^{-} \Big)\nonumber
\end{align}
and the collapse operators are 
\begin{align}
    \hat{c}_1 &= \eta \hat{\sigma}^{-}_A + \hat{\sigma}^{-}_B \nonumber \\
    \hat{c}_2 &= \sqrt{1-\eta^2} \hat{\sigma}^{-}_A. 
\end{align}
Finally, taking $\Omega_B=\Omega_A$ and $\Delta_B=-\Delta_A$, we obtain Eqs.~\eqref{eq:2qHamiltonian} and \eqref{eq:c2}.

\section{Minimum required qubit-waveguide coupling due to finite qubit lifetime}
\label{app:gamma_threshhold}
\begin{figure}[h]
    \centering
    \includegraphics{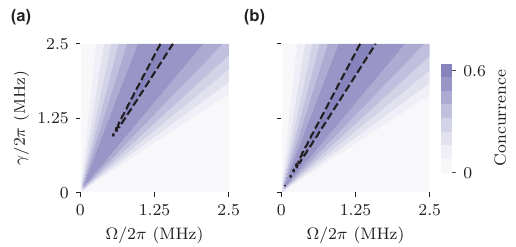}
    \caption{The figure shows the steady state concurrence of the two-qubit system for \textbf{(a)} qubits with $T_1 = 100~ \mu\mathrm{s}$ and \textbf{(b)} qubits with no intrinsic relaxation. The dashed contour lines are at concurrence $\mathcal{C}=0.56$; $\eta^2 = 0.9$. Notice in (a) that the high concurrence region is \textit{pushed} towards higher drive and coupling strengths due to intrinsic qubit relaxation. In (b), the concurrence depends only on the ratio $\Omega/\gamma$ as discussed in the main text.
    }
    \label{fig:conc_contour_intrinsic_loss}
\end{figure}

We show in Fig.~\ref{fig:1}(c) that the ratio $\Omega/\gamma$ determines the maximal achievable concurrence for given waveguide loss. However, finite qubit relaxation times require the magnitudes of $\Omega$ and $\gamma$ to be above certain minimum values. This occurs because the intrinsic loss rate of the qubit competes with the decay rate into the waveguide $\gamma$. As an example, we simulate the system with $\eta^2 = 0.9$, and qubit lifetimes of $100~ \mu\mathrm{s}$. Fig.~\ref{fig:conc_contour_intrinsic_loss}(a) shows how the high concurrence region is \textit{pushed} towards larger drive and coupling strengths, relative to the ideal case of zero intrinsic relaxation, shown in Fig.~\ref{fig:conc_contour_intrinsic_loss}(b). The dotted contour lines are at $\mathcal{C}=0.56$. To achieve this concurrence, the coupling strength to the wave guide $\gamma$ should be at least $1~\mathrm{MHz}$. This corresponds to a wave guide induced qubit relaxation time of around $160~\mathrm{ns}$.

\section{Constant concurrence in the weak driving regime}
\label{app:const-conc}

In Fig.~\ref{fig:resilience}(a), we find that for fixed $\Omega/J_{12}\sim 1$, in the weak driving limit $\Omega \ll \gamma$, the concurrence of the second qubit pair appears to saturate at a fixed value instead of falling to zero as one might expect when the driving becomes extremely weak. This turns out to be a generic feature of the system in the weak driving limit: for any nonzero waveguide transmission $\eta^2>0$ and in the limit $\Omega\to0$, the concurrence of the second pair limits to a constant nonzero value that depends only on $\eta^2$ and the constant ratio $\Omega/J_{12}$, not on $\Omega/\gamma$. Here we provide a heuristic argument for why the concurrence does not vanish in this limit.

Because the primary effect of waveguide loss is to induce unwanted single qubit $T_1$ decay on the upstream qubit (see Appendix~\ref{app:slh}), this is the primary mechanism by which the stabilization of entanglement is disrupted. We can gain insight into how much of an effect the induced $T_1$ decay has by comparing the relaxation rate of the ideal system to the rate at which excitations are lost due to the induced $T_1$. If the system relaxes quickly compared to the loss rate, then we expect a greater amount of stabilized entanglement. As we discuss in Sec.~\ref{sec:4qubit}, the weak driving and weak hopping regime, $J_{12} \lesssim \Omega \ll \gamma$, yields an ideal steady state with very little population on the first pair of qubits (cf. Eq.~\eqref{eq:2+2ss}).
We also expect that due to the weak driving, the population on the first pair of qubits remains very little throughout the entire stabilization, and we approximate the population on the first pair at all times by the steady state population, given by
\begin{align}
    \langle\hat n_1\rangle = \frac{2(\Omega/\gamma)^{4}}{2(\Omega/\gamma)^{4}+(\Omega/\gamma)^{2}+2(J_{12}/\gamma)^{2}},
\end{align}
where $\hat n_1 = |S_1\rangle\langle S_1|$ measures the singlet excitation, which is twice the excitation population on the lossy qubit. We estimate the rate at which excitations are lost due to the induced $T_1$ for $J_{12} \lesssim \Omega \ll \gamma$ by
\begin{align}
    \Gamma_{\rm loss} = \langle\hat n_1\rangle (1-\eta^2)\gamma \approx (\Omega/\gamma)^{2}(1-\eta^2)\gamma,
\end{align}
where $(1-\eta^2)\gamma$ is the induced loss rate (cf.~\eqref{eq:c2}).

To estimate the relaxation rate of the ideal system, we first assume the weak driving and weak hopping regime $J_{12} \lesssim \Omega \ll \gamma$. In this regime, the stabilization of entanglement on the second pair of qubits is well-described by the effective two-qubit master equation derived in Appendix \ref{app:adiabatic}. Moreover, we can numerically verify that the relaxation rate of the full $2+2$ system is well-approximated (up to a $\sim 1$ prefactor) by the relaxation rate of the effective theory. In particular the parameter dependence of the relaxation rate is correctly predicted. Using the result from Ref.~\cite{goviaStabilizingTwoqubitEntanglement2022} that for $\Omega/\gamma \gg 1$, the relaxation rate of the $1+1$ system is $\Gamma_{\rm rel} \sim \gamma^3/\Omega^2$, for drive strength $\Omega$ and dissipation rate $\gamma$. We find numerically that this scaling holds for $\Omega \gtrsim \gamma$. Thus, applying this result to the effective master equation, and noting that $\Omega_{\rm eff}/\gamma_{\rm eff} = \Omega/2J_{12} \gtrsim 1$ we find
\begin{align}
    \Gamma_{\rm rel} \approx \frac{\gamma_{\rm eff}^3}{\Omega_{\rm eff}^2} \simeq \frac{J_{12}^4}{\Omega^2\gamma^2}\gamma.
\end{align}
In the weak driving limit, holding $\Omega/J_{12}$ fixed, the relaxation rate thus scales with driving strength as $\Gamma_{\rm rel} \sim (\Omega/\gamma)^2$.
Comparing with the induced $T_1$ loss rate $\Gamma_{\rm loss} \sim (\Omega/\gamma)^2$, we find that as $\Omega\to 0$
\begin{align}
    \frac{\Gamma_{\rm rel}}{\Gamma_{\rm loss}} \to {\rm const}.
\end{align}
Therefore, irrespective of the amount of stabilized concurrence on the second pair, we find that in the weak driving and weak hopping limit, the stabilized concurrence saturates to a fixed value. 

\section{Adiabatic elimination and effective two-qubit theory}
\label{app:adiabatic}

Here we derive the effective 2-qubit master equation (cf.~Eq.~\eqref{eq:effective-me}) for the $2+2$ system in the weak driving limit $\Omega/\gamma\ll 1$. We thus treat the driving as a perturbation. We also assume the hopping is weak, $J_{12}/\gamma\ll1$, and treat it as perturbation. Starting from the master equation with waveguide loss,
\begin{align}
    \partial_t \hat\rho = -i[\hat H,\hat\rho] + \gamma\mathcal D[\hat c_1]\hat\rho + \gamma\mathcal D[\hat c_2]\hat\rho,
\end{align}
with the Hamiltonian and jump terms given by Eqs.~\eqref{eq:H4qb} and \eqref{eq:c2}, respectively. We take the drive detuning $\Delta=0$ here for simplicity (nothing essential is lost). 
Following the operator formalism of Ref.~\cite{reiterEffectiveOperatorFormalism2012}, we take the ``ground state manifold'' to be all states with the first pair of qubits ($A_1$ and $B_1$) in vacuum $|(00)_1\rangle$ and the ``excited state manifold'' to be the rest of the Hilbert space. The excited state manifold thus includes the three sets of states with $|01\rangle$, $|10\rangle$, and $|11\rangle$ on the first pair. Note that doubly exciting the first pair from the ground state manifold to $|11\rangle$ is second order in the perturbations (driving and hopping) we thus make a simplifying approximation and exclude the doubly-excited state from the excited state manifold.
Thus, we define the ground state manifold and excited state manifold projection operators as
\begin{align}
    \hat P_g &= (|00\rangle\langle 00|)_1\otimes \hat{1}_2,\\ \hat P_e &= (|10\rangle\langle10| + |01\rangle\langle01|)_1\otimes\hat{\mathbb{I}}_2 - \hat P_g,
\end{align}
where $\hat{1}_2$ is the identity acting on the second pair of qubits $A_2$ and $B_2$. 

We decompose the Hamiltonian into four terms $\hat H = \hat H_g + \hat H_e + \hat V_+ +\hat V_-$. The first two terms are the projections into the two manifold and the latter two terms are the off-diagonal elements connecting the manifolds; $\hat V_+$ describes excitation from the ground state manifold to the excited state manifold and $\hat V_- = \hat V_+^\dagger$ describes de-excitation:
\begin{align}
    \hat H_g &= \hat P_g \hat H \hat P_g = 0, \\
    \hat H_e &= \hat P_e \hat H \hat P_e = \frac{i\eta\gamma}{2}\left(\hat\sigma_{A,1}^+\hat\sigma_{B,1}^- - {\rm h.c.}\right)\\
    \hat V_+ &= \hat P_e \hat H \hat P_g \\
    &= J_{12}\left(|0\rangle\langle0|_{B,1}\hat\sigma_{A,1}^+\hat\sigma_{A,2}^- + |0\rangle\langle0|_{A,1}\hat\sigma_{B,1}^+\hat\sigma_{B,2}^-\right) \nonumber\\
    &+ \frac{\Omega}{2} \left( |0\rangle\langle0|_{B,1}\hat\sigma_{A,1}^+ + |0\rangle\langle0|_{A,1}\hat\sigma_{B,1}^+ \right) \nonumber\\
    \hat V_- &= \hat P_g \hat H \hat P_e \\
    &= J_{12}\left(|0\rangle\langle0|_{B,1}\hat\sigma_{A,2}^+\hat\sigma_{A,1}^- + |0\rangle\langle0|_{A,1}\hat\sigma_{B,2}^+\hat\sigma_{B,1}^-\right) \nonumber\\
    &+ \frac{\Omega}{2} \left( |0\rangle\langle0|_{B,1}\hat\sigma_{A,1}^- + |0\rangle\langle0|_{A,1}\hat\sigma_{B,1}^- \right) \nonumber
\end{align}
Note that if we retained the doubly-excited states, $\hat H_e$ would have terms $\propto \Omega,J_{12}$, but those terms always involve transitions into or out of the doubly-excited state on the first pair of qubits.

We seek an effective theory of the system in the ground state manifold, which takes the form
\begin{align}
    \partial_t\hat\rho_{\rm eff} = -i[\hat H_{\rm eff},\hat\rho_{\rm eff}] + \mathcal{D}[\hat L_{1,{\rm eff}}]\hat\rho_{\rm eff} + \mathcal{D}[\hat L_{2,{\rm eff}}]\hat\rho_{\rm eff},
\end{align}
where the effective Hamiltonian and effective jump operators are given by
\begin{align}
    \hat H_{\rm eff} &= \hat H_g - \frac{1}{2}\hat V_- \left[ \hat H_{\rm NH}^{-1} + (\hat H_{\rm NH}^{-1})^\dagger \right]\hat V_{+} \label{eq:Heff-form}\\
    \hat L_{k,{\rm eff}} &= \sqrt\gamma \hat c_k\hat H_{\rm NH}^{-1}\hat V_+ \label{eq:Lkeff-form}.
\end{align}
Here the non-hermitian Hamiltonian $\hat H_{\rm NH}$ describes evolution in the excited state manifold due to the Hamiltonian and dissipation:
\begin{align}
    \hat H_{\rm NH} &= \hat H_e -\frac{i\gamma}{2}\left(\hat c_1^\dagger\hat c_1 + \hat c_2^\dagger\hat c_2\right) \\
    &=  -\frac{i\gamma}{2}\left( \hat{\sigma}^{+}_{A,1} \hat{\sigma}^{-}_{A,1} + \hat{\sigma}^{+}_{B,1}\hat{\sigma}^{-}_{B,1} + 2\eta \hat{\sigma}^{+}_{B,1}\hat{\sigma}^{-}_{A,1}\right).\nonumber
\end{align}
Notice that $\hat H_{\rm NH}$ only acts on the first pair of qubits.

To compute the effective theory, we must invert the non-hermitian Hamiltonian. This task is made easy by the fact that it acts only on the first pair of qubits, as we only need to evaluate its matrix elements within the excited state manifold on the first pair spanned by $\{|01\rangle_1,|10\rangle_1\}$. Within this manifold, 
\begin{align}
    \hat H_{\rm NH}^{-1} = \frac{2i}{\gamma} \Big( &\hat\sigma_{A,1}^-\hat\sigma_{A,1}^+\hat\sigma_{B,1}^+\hat\sigma_{B,1}^- + \hat\sigma_{A,1}^+\hat\sigma_{A,1}^-\hat\sigma_{B,1}^-\hat\sigma_{B,1}^+ \nonumber\\
    &-2\eta \hat\sigma_{A,1}^-\hat\sigma_{B,1}^+ \Big)
\end{align}
is the inverse. 
We are thus ready to directly evaluate Eqs.~\eqref{eq:Heff-form} and \eqref{eq:Lkeff-form}. Up to irrelevant global phases, the jump operators evaluate to
\begin{align}
    \hat L_{1,{\rm eff}}^\prime &=\frac{2J_{12}}{\sqrt{\gamma}}\left(\hat{\sigma}_{B,2}^{-}-\eta\hat{\sigma}_{A,2}^{-}\right)+\frac{\Omega}{\sqrt{\gamma}}\left(1-\eta\right), \\
    \hat L_{2,{\rm eff}}^\prime &= \sqrt{1-\eta^{2}}\frac{2J_{12}}{\sqrt{\gamma}}\hat{\sigma}_{A,2}^{-}+\sqrt{1-\eta^{2}}\frac{\Omega}{\sqrt{\gamma}},
\end{align}
and the effective Hamiltonian evaluates to
\begin{align}
    \hat H_{\rm eff}^\prime &=-\frac{i}{2}\eta\frac{4J_{12}^{2}}{\gamma}\left(\hat{\sigma}_{A,2}^{+}\hat{\sigma}_{B,2}^{-}-\hat{\sigma}_{B,2}^{+}\hat{\sigma}_{A,2}^{-}\right)\\
    &+i\eta\frac{\Omega J_{12}}{\gamma}\left(\hat{\sigma}_{A,2}^{-}-\hat{\sigma}_{B,2}^{-}-\mathrm{h.c.}\right).\nonumber
\end{align}
Note that the jump operators have constant, non-operator terms $\propto\Omega$. Lindblad dissipators with jump terms of the form $\hat L = \hat X + a$ can always be decomposed into a dissipator of only the operator $\hat X$ and a Hamiltonian term via $\mathcal{D}[\hat X + a]\hat\rho = \mathcal{D}[\hat X]\hat\rho -i[(ia^{*}\hat{X}/2+{\rm h.c.}),\hat{\rho}]$.
Applying this to $\hat L^\prime_{j,{\rm eff}}$, we arrive at a new set of jump operators and effective Hamiltonian
\begin{align}
    \hat L_{1,{\rm eff}} &=\frac{2J_{12}}{\sqrt{\gamma}}\left(\eta\hat{\sigma}_{A,2}^{-}-\hat{\sigma}_{B,2}^{-}\right), \\
    \hat L_{2,{\rm eff}} &= \sqrt{1-\eta^{2}}\frac{2J_{12}}{\sqrt{\gamma}}\hat{\sigma}_{A,2}^{-},\\
    \hat H_{\rm eff} &=\frac{1}{2}\frac{2\Omega J_{12}}{\gamma}\left(\hat{\sigma}_{A,2}^{y}-\left(2\eta-1\right)\hat{\sigma}_{B,2}^{y}\right) \\
    &-\frac{i}{2}\eta\frac{4J_{12}^{2}}{\gamma}\left(\hat{\sigma}_{A,2}^{+}\hat{\sigma}_{B,2}^{-}-{\rm h.c.}\right) .\nonumber
\end{align}
Here we immediately identify the effective parameters $\gamma_{\rm eff} = 4J_{12}^2/\gamma$ and $\Omega_{\rm eff} = 2\Omega J_{12}/\gamma$
As a final step, we make local $\pm\pi/2$ rotations about $Z$ on the $A_2$ and $B_2$ qubits, respectively. This flips the relative sign between $\eta\hat{\sigma}_{A,2}^{-}$ and $\hat{\sigma}_{B,2}^{-}$ in $\hat L_{1,{\rm eff}}$ and the sign of the exchange term in the Hamiltonian, and rotates the Rabi drives from $\hat\sigma^y$ to $\hat\sigma^x$ (and flips the relative sign), thus we arrive at the effective master equation quoted in the main text.

\bibliography{references}
\end{document}

%% file: pfafflab-draft-header.tex
\usepackage{hyperref}

\usepackage{amsmath,amssymb}

\usepackage{epsfig}
\usepackage{graphics}
\usepackage{wrapfig}
\usepackage{xcolor}
\usepackage{siunitx}

\usepackage{enumitem}

\newlist{outline}{enumerate}{3}
\setlist[outline]{label=\Roman*.}
\setlist[outline,1]{label=\Roman*.}
\setlist[outline,2]{label=\Alph*.}
\setlist[outline,3]{label=\arabic*.}
\newcommand{\bout}{\begin{outline}}
\newcommand{\eout}{\end{outline}}

\newcommand{\ket}[1]{|{#1}\rangle}

\usepackage[
    textwidth=0.75in,
    textsize=footnotesize,
    color=red!25,
    colorinlistoftodos,
]{todonotes}